\documentclass[twocolumn,reprint,amsmath,amssymb
,superscriptaddress,longbibliography,prb,aps,floatfix]{revtex4-1}
\usepackage{graphicx,hyperref}
\hypersetup{colorlinks=true, linkcolor=red, citecolor = blue}
\usepackage{bm}
\usepackage{amsmath, amssymb, upgreek}
\def\vector#1{\bm{#1}} 
\def\jj{\,\mathrm{j}}
\def\cc{\mathrm{c.c.}}
\def\sub#1{_{\mathrm{#1}}}
\def\sur#1{^{\mathrm{#1}}}

\graphicspath{{./figs/}}  
\begin{document}
\title{Optical heterodyne imaging of magnetostatic modes in one-dimensional magnonic crystals}

\author{S.~Z.~Baba}
\email{baba@noneq.t.u-tokyo.ac.jp}
\author{Y.~Nakata}
\author{Y.~Ito}
\author{R.~Hisatomi}
\affiliation{Research Center for Advanced Science and Technology (RCAST), The University of Tokyo, Meguro-ku, Tokyo 153-8904, Japan}
\author{Y.~Nakamura}
\affiliation{Research Center for Advanced Science and Technology (RCAST), The University of Tokyo, Meguro-ku, Tokyo 153-8904, Japan}
\affiliation{Center for Emergent Matter Science (CEMS), RIKEN, Wako, Saitama 351-0198, Japan}
\author{K.~Usami}
\email{usami@qc.rcast.u-tokyo.ac.jp}
\affiliation{Research Center for Advanced Science and Technology (RCAST), The University of Tokyo, Meguro-ku, Tokyo 153-8904, Japan}

\date{\today}

\begin{abstract}
We demonstrate a real-space imaging of a heterodyne signal of light that is produced as a result of the Brillouin light scattering by coherently driven magnons in magnetostatic modes. With this imaging technique, we characterize surface magnetostatic modes (Damon-Eshbach modes) in a one-dimensional magnonic crystal, which is formed by patterned aluminum strips deposited on the ferromagnetic film. The modified band structures of the magnonic crystal are deduced from the Fourier transforms of the real-space images. The heterodyne imaging provides a simple and powerful method to probe magnons in structured ferromagnetic films, paving a way to investigate more complex phenomena, such as Anderson localization and topological transport with magnons.
\end{abstract}

\maketitle

\section{Introduction}

Bloch electrons in crystals exhibit ample phenomena in solid state physics largely 
due to the associated band structures~\cite{ashcroft1976_SolidStatePhysics}. 
Inspired by that, photonic crystals 
have been explored to manipulate photons by means of periodic modulation of the refractive index
~\cite{yablonovitch1987_InhibitedSpontaneousEmission,john1987_StrongLocalizationPhotons}. 
Similar ideas can be applied to other bosonic excitations, such as phonons and magnons.

Among other bosonic excitations 
magnons in ferromagnetic insulators are peculiar in the following three senses:
They (i) break time-reversal symmetry (ii) have long coherence time, and (iii) are optically detectable. 
These features offer interesting opportunities to magnons as a carrier of information, entropy, energy, momentum, and angular momentum~\cite{kruglyak2010_Magnonics,serga2010_YIGMagnonics,nikitov2015_MagnonicsNewResearch,chumak2015_MagnonSpintronics}. In the field of \textit{magnonics}, magnon band engineering with \textit{magnonic crystals}~\cite{krawczyk2014_ReviewProspectsMagnonic} has been explored to manipulate propagation and localization of magnons \cite{demokritov2017_SpinWaveConfinement} for data processing with magnons~\cite{chumak2017_MagnonicCrystalsData}, and to render magnons topologically protected~\cite{shindou2013_TopologicalChiralMagnonic,li2018_TopologicalMagnonModes}. 

To facilitate further development of magnonics, imaging of magnetostatic modes in real-space can be a vital approach. The micro-focused Brillouin light scattering ($\upmu$-BLS) imaging technique~\cite{Demidov2004,sebastian2015_MicrofocusedBrillouinLight} has been developed and widely used in this respect.
The $\upmu$-BLS imaging is capable of mapping the intensity profile of magnetostatic modes
 with high spatial resolution and high sensitivity. 
These are achieved by meticulously filtering the photons from the inelastic Brillouin scattering 
with Fabry-P\'{e}rot cavities and detecting the filtered photons with a photon counter~\cite{sebastian2015_MicrofocusedBrillouinLight}. 
The photon counting method used in the $\upmu$-BLS imaging is intrinsically incapable of detecting
 the phase of light. The phase information is, however, essential to fully understand 
 the dynamics of magnons and plays a vital role in topological physics~\cite{vanderbilt2018}. 
 With an additionally introduced local oscillator it is possible to make the $\upmu$-BLS imaging
  phase-sensitive~\cite{sebastian2015_MicrofocusedBrillouinLight,serga2006_PhasesensitiveBrillouinLight,fohr2009_PhaseSensitiveBrillouin,vogt2009_AllopticalDetectionPhase,demidov2009_ControlSpinwavePhase}.

Here, we demonstrate an optical heterodyne $\upmu$-BLS imaging of magnetostatic modes,
 where the beat note between the \textit{scattered sideband} optical field 
 and the input \textit{carrier} field is detected with a high-speed photodetector. 
 The advantage of the heterodyne imaging is to simultaneously obtain the amplitude 
 and the phase of a microwave-driven magnetostatic mode. 
 We use this method to characterize surface magnetostatic modes 
 (Damon-Eshbach modes)~\cite{eshbach1960_SurfaceMagnetostaticModes,damon1961_MagnetostaticModesFerromagnet,stancil2009_SpinWavesTheory,gurevich1996_MagnetizationOscillationsWaves} 
 in a one-dimensional~(1D) magnonic crystal formed by patterned aluminum strips deposited on the ferromagnetic film. 
 From the Fourier transforms of the spatial images, the resultant modification of the band structure is verified. 
 This \textit{frequency-domain} spectroscopic reconstruction of the dispersion of the magnetostatic mode can be viewed as a complementary approach to the all-optical \textit{time-domain} reconstruction of spin-wave dispersion~\cite{hashimoto2017_SWaT}. 


\section{Experimental setup}

\begin{figure*}[t]
\includegraphics[width=16.2 cm]{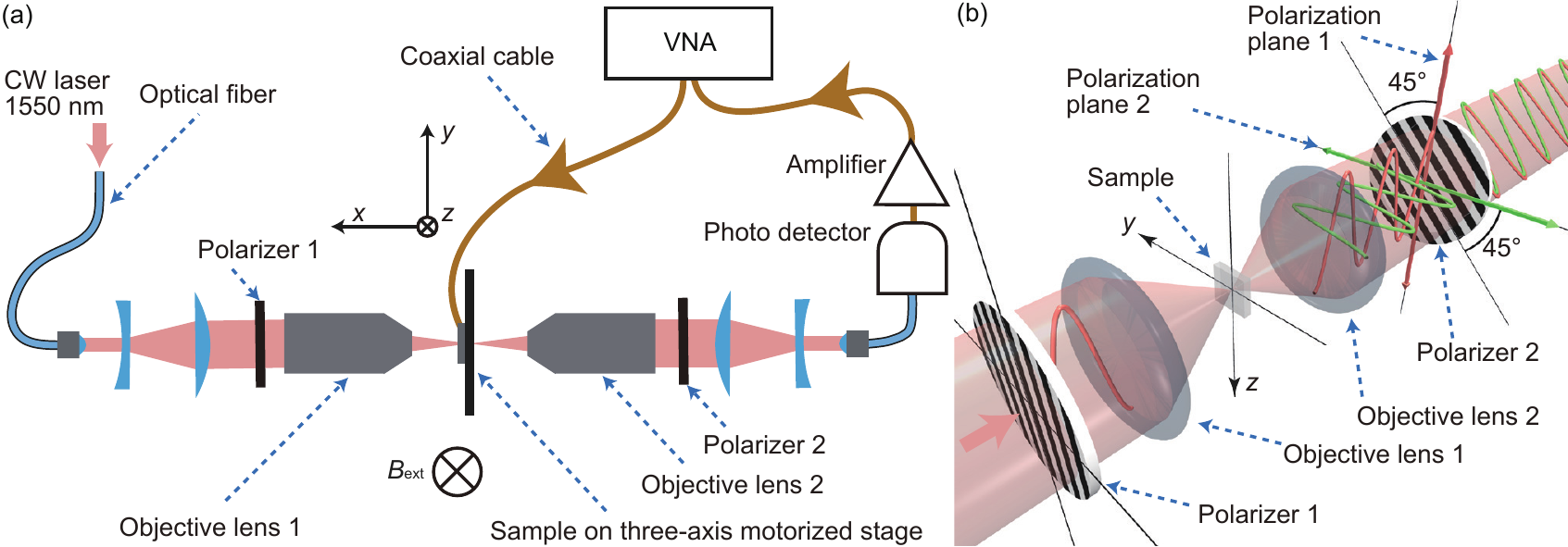}
\caption{(a) Experimental setup. A light field from a CW laser with the wavelength of 1550~nm is sent through polarizer 1 and propagates toward the negative $x$-direction. The input field is tightly focused on a sample by objective lens 1. The sample can be moved with a three-axis motorized stage. 
The scattered sideband fields, as well as the unscattered carrier field, 
are collected by objective lens 2 and coupled to a photodetector through polarizer 2. 
The heterodyne beat note between the scattered sideband fields and the carrier field is amplified and fed into a vector network analyzer (VNA). 
(b) The detailed picture of the setting of polarizers: Polarizer~1 sets the input carrier field $z$-polarized. 
The scattered sideband field is $y$-polarized. 
The carrier and sideband fields are mixed through polarizer~2, 
which is rotated by $45^\circ$ with respect to polarizer~1.} \label{fig:setup}
\end{figure*}

Figure~\ref{fig:setup} shows the simplified experimental setup for the optical heterodyne imaging. A light field from a CW laser with the wavelength of 1550~nm propagates toward the negative $x$-direction as shown in Fig.~\ref{fig:setup}(a). Here the input polarization is purified along the $z$-axis with the first polarizer as shown in Fig.~\ref{fig:setup}(b). The input field is tightly focused (waist radius $\approx$ 4.3~$\upmu$m) onto a sample of a ferromagnetic film by an objective lens~[\textsc{Mitsutoyo} M Plan APO NIR \(10\times\)]. The sample can be moved in three dimensions with respect to the focus of the input optical field with a three-axis stage with stepper motor actuators~[\textsc{Thorlabs} ZFS13B]. Although the input field has the relatively long Rayleigh range (\(\sim\)~37~\(\upmu\)m), the field is mainly scattered by the magnons on the upper side of the film since the antenna strongly excites the magnons on that side. The field thus undergoes Brillouin light scattering which creates sideband fields with a different polarization as we will describe below. The scattered sideband fields, as well as the unscattered carrier field, are collected by another lens~[\textsc{Mitsutoyo} M Plan APO NIR \(10\times\)] and coupled to a high-speed photodetector~[\textsc{New Focus} 1554-B] through a single-mode fiber after the second polarizer rotated by 45 degrees from the $y$-axis as shown in Fig.~\ref{fig:setup}(b). The heterodyne beat note between the scattered sideband fields and the carrier field is then amplified [\textsc{Mini Circuits} ZX60-83LN-S+] and fed into a vector network analyzer~[\textsc{Agilent Technologies} N5232A], which demodulates the beat signal with the drive signal used to excite the magnons. 

The Brillouin light scattering takes place when the field creates additional magnons or annihilates magnons. 
Since each magnon possesses a spin angular momentum $\hbar$, 
creation or annihilation of a magnon means the change of the polarization of a photon in the scattered field. 
This correspondence between the magnon and the scattered photon in terms of spin angular momentum is due to the conservation of angular momentum. Furthermore, the number of the scattered photons is proportional to that of magnons, and the phase information of excited magnons is also coherently transferred to the scattered photons. Thus by looking at the beat note between the scattered photons and the input photons, we could obtain full information regarding the magnons involved in the scattering. 

At this juncture, let us discuss the performance of the heterodyne $\upmu$-BLS imaging. The detectable magnon bandwidth by the heterodyne $\upmu$-BLS imaging is only limited by the photodetector bandwidth, which can be as large as 10 GHz. The sensitivity of the heterodyne imaging is shot-noise-limited, which differs from the conventional photon-counting-based $\upmu$-BLS imaging for which the sensitivity is limited by the dark counts. This means that, given that the number of input carrier photons is $N$ per second, the \textit{noise-equivalent scattering rate} would be $\sqrt{N}$ per second. 

\begin{figure}[tb]
 \includegraphics[width=8.6cm]{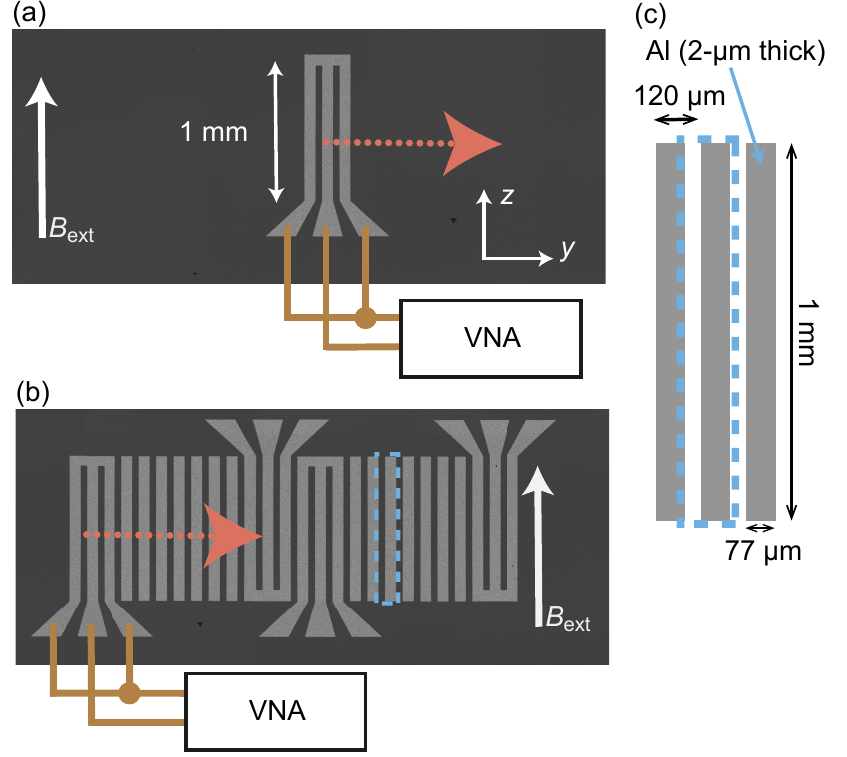}
\caption{Micrographs of samples: 
(a) plain ferromagnetic film and
(b) 1D magnonic crystal with aluminum stripes. 
A vector network analyzer (VNA) is connected to a metallic antenna for excitation.
Magnons excited by the antennas propagate toward the positive \(y\)-direction.
The dotted arrows show scanning paths of a focused laser spot over the sample.
 The static magnetic field ($B_{\mathrm{ext}} \sim 106$~mT) is applied for each sample along the $z$-axis.
 The unit cell of the magnonic crystal is highlighted with the dashed rectangle.
  (c) Unit-cell design for the magnonic crystal.
\label{fig:sample}}
\end{figure}

\section{Results}

\subsection{Plain ferromagnetic film}

We first show how the optical heterodyne imaging technique works in a simple setup. To this end, we focus here on the Damon-Eshbach modes with a plain ferromagnetic thin film. Here we see that the optical heterodyne imaging technique enables us to reconstruct the peculiar dispersion of the modes.
 
\subsubsection{Sample}
The samples are tangentially magnetized under a static magnetic field ($B\sub{ext}\sim$ 106~mT) along the positive $z$-direction as shown in Fig.~\ref{fig:setup}, which is produced by a pair of permanent magnets with a pure-iron magnetic circuit and an additional solenoid winding around the magnetic circuit so as fine tune the static magnetic field.
 The samples are thin films made of yttrium iron garnet (YIG) [thickness: $d=$~9.5~$\upmu$m, crystal orientation: (111)] on gadolinium gallium garnet (GGG) substrate [thickness: 0.5~mm, crystal orientation: (111)].
Figure~\ref{fig:sample}(a) shows a sample with a microwave antenna made of aluminum, 
which is deposited on the plain YIG film. 
The antenna is designed to excite mainly magnons in the Damon-Eshbach mode~\cite{eshbach1960_SurfaceMagnetostaticModes,damon1961_MagnetostaticModesFerromagnet}.
The magnons in the Damon-Eshbach mode are propagating only in one direction on the upper surface of the film (to the positive $y$-direction) and in the opposite direction on the bottom surface (to the negative $y$-direction).

\subsubsection{Dispersion of the Damon-Eshbach modes}

\begin{figure}[tbp]
\includegraphics[width=8.6cm]{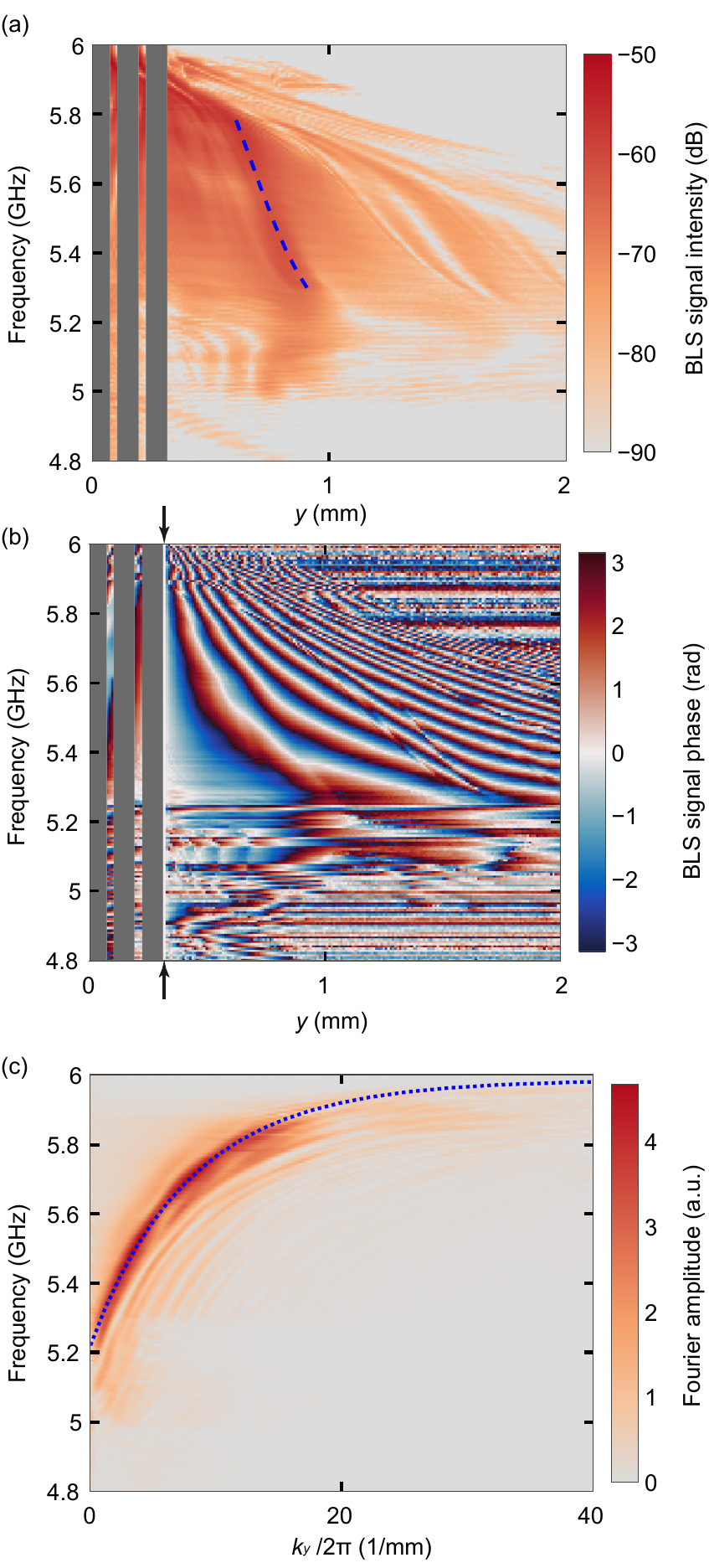}
\caption{One-dimensional real-space images of (a) the intensity and (b) the phase 
of the heterodyne signal from the Brillouin light scattering by magnons in the plain ferromagnetic film as a function of the microwave frequency (vertical axis) 
at which magnons are excited. The dashed line in panel (a) represents the position where the caustic wave beam coming from the antenna corner hits the 1D scanning line (see Appendix~\ref{app:magnon_caustics}). 
The phase reference point is represented by arrows in panel (b). 
The vertical gray strips represent the regions where aluminum is deposited. In these regions, we cannot obtain the heterodyne signal. Thus, we filled the signal data of the region with zero when we performed the Fourier transform.
(c) Dispersion reconstructed from the real-space images shown in panels (a) and (b) by a Fourier transform for the range \(0.32 \text{~mm}< y < 2 \text{~mm}\). 
The dotted line indicates the theoretically evaluated dispersion with $M_s=166.4$~kA/m and $B\sub{ext}=\mu_{0} H\sub{ext}=109.0$~mT.
} \label{fig:result1}
\end{figure}

We scanned a focused laser spot over the sample along the dotted arrow shown in Fig.~\ref{fig:sample}(a).
Figures~\ref{fig:result1}(a) and (b) respectively show the 1D real-space images of the intensity (power) and phase of the heterodyne signal as a function of the drive frequency for the magnons in the plain YIG film shown in Fig.~\ref{fig:sample}(a). 
Here, the phase is defined with respect to the edge of the antenna as shown by arrows in Figs.~\ref{fig:result1}(b). 
The intense signals are found around the dashed lines in Fig.~\ref{fig:result1}(a). These are attributed to a beam-like mode, called \textit{the caustic wave beam}~\cite{schneider2010}, which is diagonally propagating from the corner of the antenna. We shall explain how the caustic wave beam emerges from the peculiar hyperbolic-like dispersion of the Damon-Eshbach modes in Appendix~\ref{app:magnon_caustics}.

The measured 1D real-space images of the intensity and phase, shown in Figs.~\ref{fig:result1}(a) and (b), respectively, can be used to reconstruct the dispersion relations by performing the discrete Fourier transform of the real-space images. The transformation is given by
\begin{equation}
 [\mathcal{F}(\vector{s})]_m =
  \frac{1}{N} \sum_{n=0}^{N-1} s_n \exp\left(2\pi \jj \frac{m n}{N}\right),  \label{eq:2}
\end{equation}
where the complex amplitude of the optical heterodyne signal [measured as $S_{21}$ in the VNA shown in Fig.~\ref{fig:setup}(a)] at $x=x_j=j \Delta x$ ($j=0,1,\cdots, N-1$) is denoted by $s_j$, and $[\mathcal{F}(\vector{s})]_m$ with $\vector{s}=[s_0\ s_1\ \cdots \ s_{N-1}]^\mathrm{T}$ 
is the complex amplitude at the angular wavenumber $k_m=m\Delta k$, the step of which is $\Delta k=2\pi/(N\Delta x)$ 
with the length interval $\Delta x$ of data points. 
Figure~\ref{fig:result1}(c) shows the result of the Fourier transform for the range \(0.32 \text{~mm}< y < 2 \text{~mm}\) of the 1D real-space image.

In the magnetostatic regime, the Damon-Eshbach mode obeys the following dispersion relation 
\cite{damon1961_MagnetostaticModesFerromagnet,stancil2009_SpinWavesTheory,gurevich1996_MagnetizationOscillationsWaves}:
\begin{equation}
k = -\frac{1}{2d}\ln\left[1+\frac{4}{{\omega_{M}}^2}\Big[\omega_0(\omega_0+\omega_{M})-\omega^2\Big]\right], \label{eq:1}
\end{equation}
where $\omega_{M}=-\gamma \mu_0 M_s$ and $\omega_0=-\gamma B\sub{ext}$ with $d$ being thickness of the film, $\mu_0$ 
being the vacuum permeability, $M_s$ being the saturation magnetization, and $\gamma$ $(<0)$ being the gyromagnetic ratio. 
For YIG, we have $\gamma \approx -1.761\times10^2$~rad GHz/T. 
The blue dotted line in Fig.~~\ref{fig:result1}(c) corresponds to the theoretical  dispersion of the Damon-Eshbach mode [the inverse form of Eq.~(\ref{eq:1})]:
\begin{equation}
 \omega = \sqrt{\omega_0(\omega_0+\omega_{M})+\frac{{\omega_{M}}^2}{4} [1-\exp(-2kd)]}, \label{eq:3}
\end{equation}
where $M_s$ and $B\sub{ext}$ are used as the fit parameters. 
From the  
least squares fit to the spatial peak positions of the power in each frequency, we obtain $M_s=166.4\pm 0.5$~kA/m and $B\sub{ext}=\mu_{0} H\sub{ext}=109.0\pm0.2$~mT, 
which are reasonable agreements with the expected value of $M_s$ at room temperature, 140~kA/m~\cite{stancil1993_TheoryMagnetostaticWaves}, and the measured value of $B\sub{ext}\sim$~106~mT, respectively. This suggests that the method to reconstruct the dispersion relation from the real-space imaging works well.  It is emphasized that to reconstruct the dispersion from the real-space imaging it is vital to have both of the amplitude and the phase of the optical signal. 
We can reconstruct not only the amplitude of the dispersion shown in Fig.~\ref{fig:result1}(c) but also the phase in the reciprocal space. 
The phase information in the reciprocal space could be useful when the topological aspects of the dispersion are of interest.
  
\subsection{1D magnonic crystal}
We now apply the optical heterodyne imaging technique to the investigation of magnons in a 1D magnonic crystal for the Damon-Eshbach modes. 

\subsubsection{Sample}
Figure~\ref{fig:sample}(b) shows a sample with four microwave antennas embedded in the 1D magnonic crystal~\cite{kanazawa2014_SpinWaveLocalization,mruczkiewicz2014_ObservationMagnonicBand,bessonov2015_MagnonicBandGaps,kanazawa2015_MetalThicknessDependence}. 
The unit cell for the magnonic crystal is defined in a rectangle of 1~mm $\times$ 120~$\upmu$m as shown in the dashed rectangle in Fig.~\ref{fig:sample}(c). 
The aluminum region imposes an additional boundary condition on the tangential electric field, and thus modifies the effective magnetic dispersions. 

\subsubsection{One-dimensional magnonic crystal and bandgap formation}
\begin{figure}[htbp]
\includegraphics[width=8.6cm]{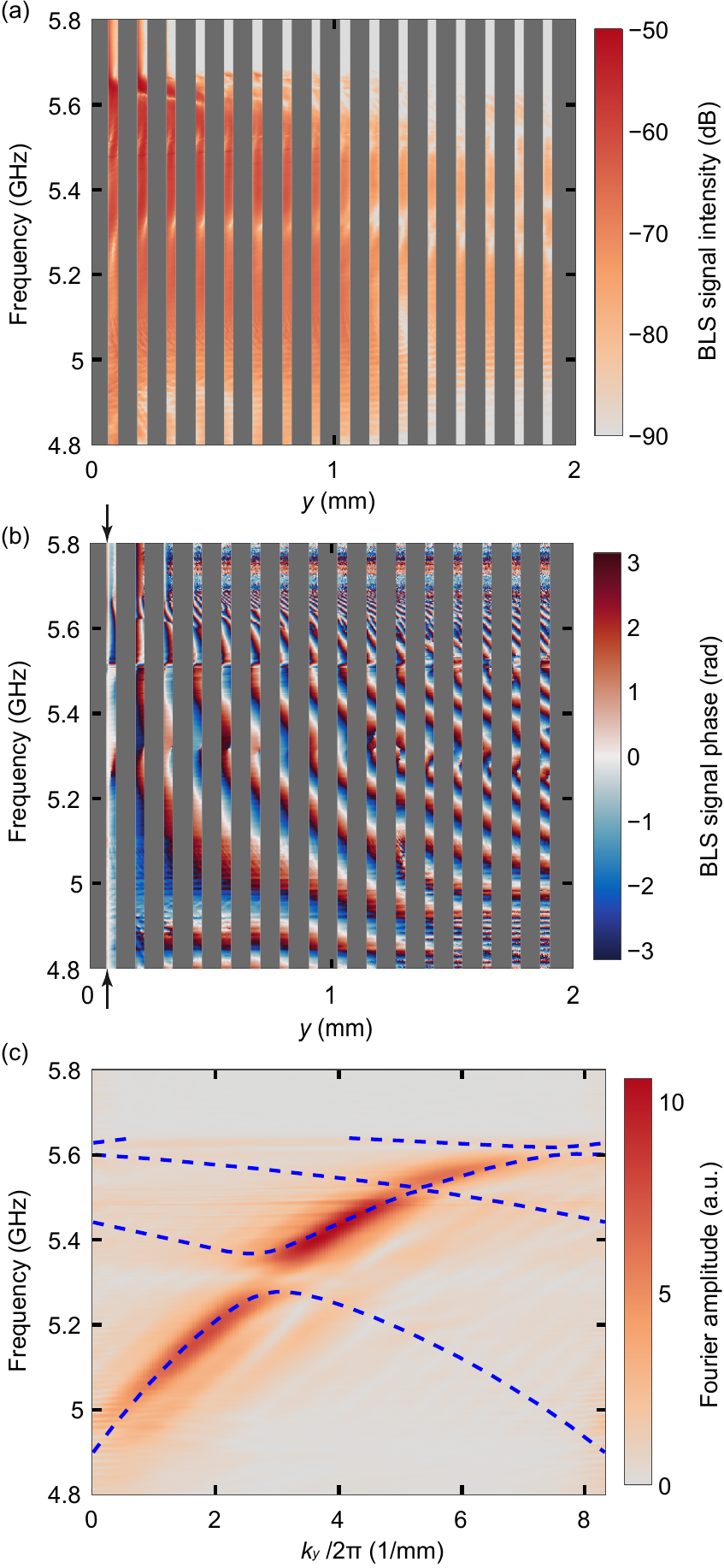}
\caption{One-dimensional real-space images of (a) the intensity and (b) the phase 
of the heterodyne signal from the Brillouin light scattering by magnons in the 1D magnonic crystal as a function of the microwave frequency (vertical axis) 
at which magnons are excited. The phase reference point is represented by arrows in panel (b). The vertical gray strips represent the regions where aluminum is deposited. In these regions, we cannot obtain the heterodyne signal. Thus, we filled the signal data of the region with zero when we performed the Fourier transform. 
(c) Dispersion reconstructed from the real-space images shown in panels (a) and (b) by a Fourier transform for the range \(0.31 \text{~mm}< y < 2 \text{~mm}\). 
The dashed line indicates the theoretically calculated dispersion with $M_s=  164.91$~kA/m and $B_\mathrm{ext}= 99.56$~mT using a transfer matrix model.
} \label{fig:result2}
\end{figure}

We scanned a focused laser spot over the sample along the dotted arrow shown in Fig.~\ref{fig:sample}(b).
Figures~\ref{fig:result2}(a) and (b) respectively show the frequency dependence of the real-space imaging of the intensity and phase of the heterodyne signal obtained 
with the 1D magnonic crystal. 
With these data, we performed a discrete Fourier transform~[Eq.~(\ref{eq:2})] for the range \(0.31 \text{~mm}< y < 2 \text{~mm}\). 
Figure~\ref{fig:result2}(c) shows the resultant Fourier transform. 
We can clearly see the bandgaps around 5.3~GHz and 5.5~GHz. Note that in the results shown in Fig.~\ref{fig:result2}(c) the portion of the dispersion curves where the magnon has a negative group velocity is absent. This can be understood by the fact that the magnons excited by the antenna, mainly on the upper side of the film, flow from left to right. 

Now let us analyze the observed bandgaps. The magnonic crystal shown in Fig.~\ref{fig:sample}(b) has metallic stripes. 
The region with a top metal is denoted by $\mathrm{M}$, while that without the metal is written as $\mathrm{A}$. 
If the metal thickness is larger than the microwave penetration depth ($\approx$1.2~$\upmu$m at 5~GHz for aluminum with the resistivity 
$\rho_\mathrm{DC}=27$~n$\Omega \cdot$m), the metal can be safely considered perfectly conducting, 
and the dispersion relation in region M can be given by \cite{seshadri1970_SurfaceMagnetostaticModes, gurevich1996_MagnetizationOscillationsWaves}:
\begin{equation}
k_{\pm} = -\frac{1}{2d}\ln\left[ \left(1+2\frac{\omega_0}{\omega_{M}} \pm 2\frac{\omega}{\omega_{M}}\right)
\frac{\omega_0+\omega_{M} \mp \omega}{\omega_0+\omega_{M} \pm \omega}\right],  \label{eq:5}
\end{equation}
for propagation toward the positive (negative) $y$-direction. This modification of the dispersion compared to that in region M [given in Eq.~(\ref{eq:1})] is the key to realizing the magnonic crystal.

Bandgap formation due to the magnonic crystal can be captured by a simple one-dimensional transfer matrix model. 
Consider a \textit{magnetostatic potential}~\cite{gurevich1996_MagnetizationOscillationsWaves}, $\psi(y,t) = \tilde{\psi}(y)\exp(\jj \omega t) + \cc$, inside the magnonic crystal with the complex amplitude $\tilde{\psi}(y)$, which is defined in terms of the induced magnetic field along the $y$-axis, $h(y,t)$, by
\begin{equation}
h(y,t) \approx - \frac{\partial}{\partial y} \psi(y,t).
\end{equation}
Here, for simplicity, we suppress $z$ and $x$ dependencies in $\psi$. We can decompose $\tilde{\psi}(y)$ into the counter propagating components $\tilde{\psi}_+(y)$ and $\tilde{\psi}_-(y)$ as $\tilde{\psi}(y)=\tilde{\psi}_+(y)+\tilde{\psi}_-(y)$, where $+$ and $-$ represent positive and negative propagation
along the $y$-axis, respectively. These components are combined and denoted as
$
 \tilde{\vector{\psi}}(y)=
  \begin{bmatrix}
   \tilde{\psi}_+(y)\\
   \tilde{\psi}_-(y)
  \end{bmatrix}.
$

\begin{figure}[tb]
 \includegraphics[width=8.6cm]{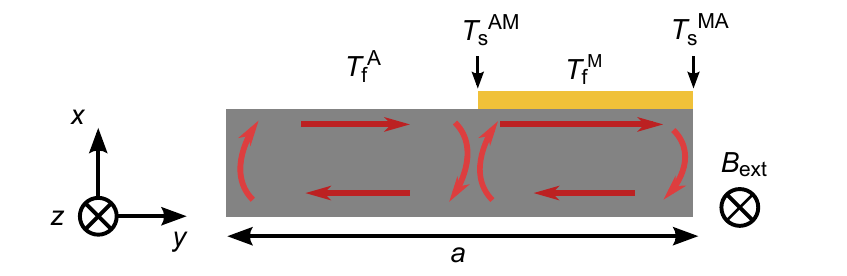}
\caption{Schematic model of the cross section of a unit cell of the magnonic crystal. $T\sub{f}^{\mathrm{A}}$ ($T\sub{f}^{\mathrm{M}}$) is a free propagation transfer matrix in region A (M) while $T\sub{s}^{\mathrm{AM}}$ ($T\sub{s}^{\mathrm{MA}}$) is a transfer matrix at the interface between region A (M) and region M (A). The external magnetic field is along the $z$-axis.} \label{fig:transfer_matrix_model}
\end{figure}

To calculate the dispersion relation of the magnonic crystal, we define a transfer matrix of the unit cell  shown in Fig.~\ref{fig:transfer_matrix_model} as
\begin{equation}  
T\sub{unit}=T\sub{f}\sur{A}T\sur{AM}\sub{s} T\sub{f}\sur{M}T\sur{MA}\sub{s}.
\end{equation}
Here, $T\sub{f}^{\mathrm{A}}$ ($T\sub{f}^{\mathrm{M}}$) is a free propagation transfer matrix in region A (M). At the interface between region A (M) and region M (A) a distinct transfer matrix, $T\sub{s}^{\mathrm{AM}}$ ($T\sub{s}^{\mathrm{MA}}$), can be defined. We shall explain more details regarding each transfer matrix later on. For a Bloch state $\tilde{\vector{\psi}}$ with the $y$-component  $k\sub{c}$ of the crystal wavenumber, we have
\begin{equation}
 T\sub{unit}\tilde{\vector{\psi}} = \exp(\jj k\sub{c} a) \tilde{\vector{\psi}}  \label{eq:8}
\end{equation}
with $a$ being the unit cell length. Therefore, by diagonalizing  $T\sub{unit}$ for a given $\omega$, we obtain $k\sub{c}$ as a function of $\omega$.

Now we explain the details of each transfer matrix. The transfer matrix of free propagation along the length $l$ in region A is reciprocal and written as
\begin{equation}
 T\sub{f}^{\mathrm{A}} =
 \begin{bmatrix}
  \exp(\jj k l) & 0\\
  0 & \exp(-\jj k l)
 \end{bmatrix},
\end{equation}
where, for a given $\omega$, $k$ is given by Eq.~(\ref{eq:1}). On the other hand, that in region M is nonreciprocal and written as
\begin{equation}
 T\sub{f}^{\mathrm{M}} =
 \begin{bmatrix}
  \exp(\jj k_{+} l) & 0\\
  0 & \exp(-\jj k_{-} l)
 \end{bmatrix},
\end{equation}
where, for a given $\omega$, we obtain $k_\pm$ through Eq.~(\ref{eq:5}). 

At the interface $y=\xi$ between region M and region A, for instance, the scattering occurs and mixes up $\tilde{\psi}_{+}$ and $\tilde{\psi}_{-}$. This mixing is denoted with the transfer matrix $T\sub{s}^{\mathrm{AM}}$ by $\tilde{\vector{\psi}}^{\mathrm{A}} (\xi) = T\sub{s}^{\mathrm{AM}}\tilde{\vector{\psi}}^{\mathrm{M}} (\xi)$. Here, $T\sub{s}^{\mathrm{AM}}$ can be determined from the following boundary conditions: At $y=\xi$, $\tilde{\psi}^{\mathrm{A}}$ and $\tilde{B}_y^{\mathrm{A}} \approx -\mu_{yy} \partial_y \tilde{\psi}^{\mathrm{A}}$ must be equal to $\tilde{\psi}^{\mathrm{M}}$ and $\tilde{B}_y^{\mathrm{M}} \approx -\mu_{yy} \partial_y \tilde{\psi}^{\mathrm{M}}$, respectively. Here $\mu_{yy}$ represents the $yy$-component of the permeability tensor. These conditions amount to
\begin{equation}
M_{\mathrm{A}} \tilde{\vector{\psi}}^{\mathrm{A}}(\xi) = M_{\mathrm{M}} \tilde{\vector{\psi}}^{\mathrm{M}}(\xi),
\end{equation}
where $M_{\mathrm{A}} =\begin{bmatrix}
1 & 1\\
k & -k
\end{bmatrix}$ and $M_{\mathrm{M}} =\begin{bmatrix}
1 & 1\\
k_{+} & -k_{-}
\end{bmatrix}$.The transfer matrix $T\sub{s}^{\mathrm{AM}}$ at the interface $y=\xi$ is thus given by 
\begin{equation}
T\sub{s}^{\mathrm{AM}} = M_{\mathrm{A}}^{-1} M_{\mathrm{M}}.
\end{equation} 
The transfer matrix $T\sub{s}^{\mathrm{MA}}$ is similarly obtained as 
\begin{equation}
T\sub{s}^{\mathrm{MA}} = M_{\mathrm{M}}^{-1} M_{\mathrm{A}}.
\end{equation}

For a given angular frequency $\omega$, we have a $2\times 2$ transfer matrix $T\sub{unit}$ for the unit cell of the magnonic crystal given by Eq.~(\ref{eq:8}). From this characteristic equation, we can obtain $k\sub{c}$ as a function of $\omega$, which is shown as the blue dashed lines in Fig.~\ref{fig:result2}(c). 
Here, the fit parameters are $M_s$ and $B\sub{ext}$. These values are found as $M_s=  164.91\pm 0.06$~kA/m and $B_\mathrm{ext}= 99.56\pm 0.03$~mT, which are similar to those obtained from the plain film above. This agreement indicates that the model we developed captures the physics behind the bandgap formation and that the optical heterodyne imaging provides a powerful means to diagnose magnon propagation in magnonic crystals and related artificial magnetic structures.

\section{Summary}
The heterodyne $\upmu$-BLS imaging is demonstrated to study a simple 1D magnonic crystal. In this method, both the amplitude and the phase of the optical heterodyne signal, which stems from the scattering by the microwave-driven magnons, are simultaneously obtained. 
The Fourier transforms of the real-space images can be used to obtain dispersion relations of the magnetostatic modes and to verify the opening of the magnonic bandgap caused by the 1D magnonic crystal. Our results show that the heterodyne $\upmu$-BLS imaging could be a simple and powerful way to probe magnons in ferromagnetic films, paving a way to investigate more complex phenomena with magnons in artificial structures.

\section{Acknowledgements}
We thank Ryo Iguchi for suggesting the idea of the optical heterodyne imaging. We also thank R.~Yamazaki, A.~Okada, S.~Daimon, T.~Goto, and R.~Shindou for useful discussions. This work is partly supported by the Matsuo Foundation, JSPS KAKENHI Grant Number 26220601, and JST ERATO Grant Number JPMJER1601.

\appendix
\renewcommand\thefigure{\thesection\arabic{figure}}    
\setcounter{figure}{0}    
\section{Effect of magnon caustics \label{app:magnon_caustics}}

Here, we discuss the physical origin of the intense signals around the dashed lines in Fig.~\ref{fig:result1}(a). We see that the signal is due to a non diffracting beam like mode (called the \textit{caustic wave beam}~\cite{schneider2010,Gieniusz2013, Davies2015}), which is generated around the antenna corner.
Note that, except for the corner, the propagation direction of magnons generated from the antenna is usually restricted to 
be perpendicular to the lines due to the structure of the antenna.

\begin{figure}[tbp]
 \centering
  \includegraphics[width=8.6cm]{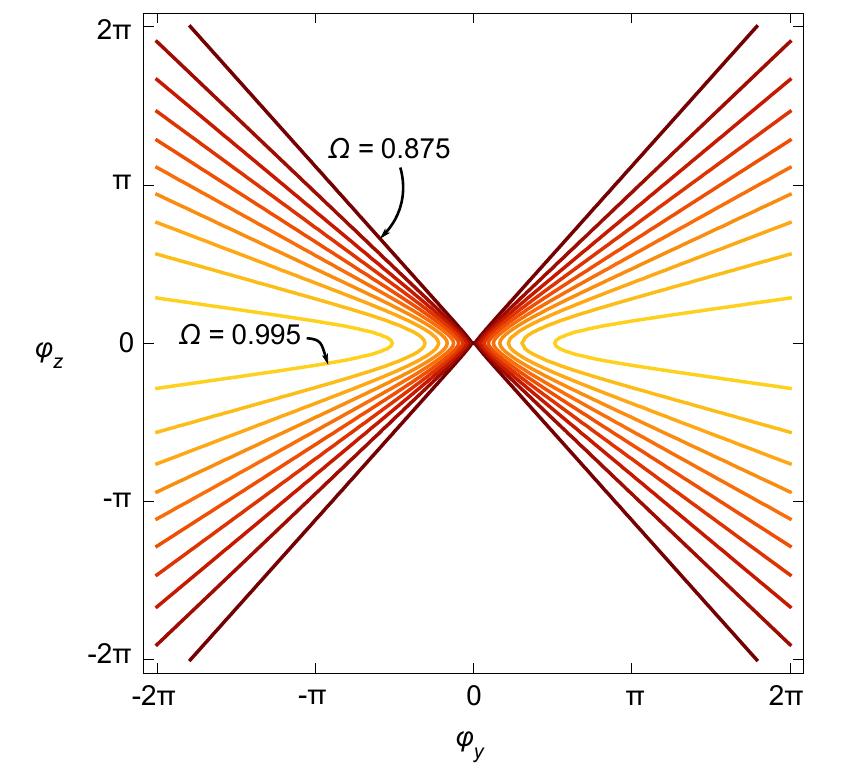}

\caption{Isofrequency contours given by Eq.~(\ref{eq:4}) of the Damon-Eshbach modes for $\Omega=\omega/\omega_{M}$ ranging from $0.875$ to $0.995$\label{fig:corner_effect}.}
\end{figure}

To this end, let us consider the dispersion of the Damon-Eshbach mode on a 2D surface. 
The dispersion of the Damon-Eshbach mode is peculiar; the isofrequency contour is a hyperbolic like shape as opposed to the typical elliptical shape. 
The isofrequency contour is given by~\cite{damon1961_MagnetostaticModesFerromagnet}
\begin{align}
  &\! & (\varphi_x\sur{e})^2 + 2 (1+\kappa) \varphi_x\sur{i}\varphi_x\sur{e}
  \cot(\varphi_x\sur{i}) \nonumber \\
  &\ &\ \ \ - (1+\kappa)^2 (\varphi_x\sur{i})^2 -(\nu \varphi_y)^2=0,  \label{eq:4}
\end{align}
with the \textit{dimensionless wavenumbers}, $\varphi\sur{e}_x$, $\varphi_x\sur{i}$, $\varphi_y$, and $\varphi_z$, which are defined as
\begin{align}
\varphi\sur{e}_x &= \sqrt{{\varphi_y}^2 + {\varphi_z}^2}, \\
\varphi_x\sur{i} &= \jj\sqrt{{\varphi_y}^2 + {\varphi_z}^2/(1+\kappa)},\\
\varphi_y &= k_y d,\\
\varphi_z &= k_z d,
\end{align}
with $\kappa \equiv \omega_{M}\omega\sub{0}/({\omega_0}^2 - \omega^2)$ and 
$\nu \equiv \omega_{M}\omega/({\omega_0}^2 - \omega^2)$. 
Here, $d$ is the thickness of the film. Figure~\ref{fig:corner_effect} depicts such isofrequency contours for 
$\Omega \equiv \omega/\omega_{M}$ from 0.875 to 0.995.

\begin{figure}[tbp]
 \centering
  \includegraphics[width=8.6cm]{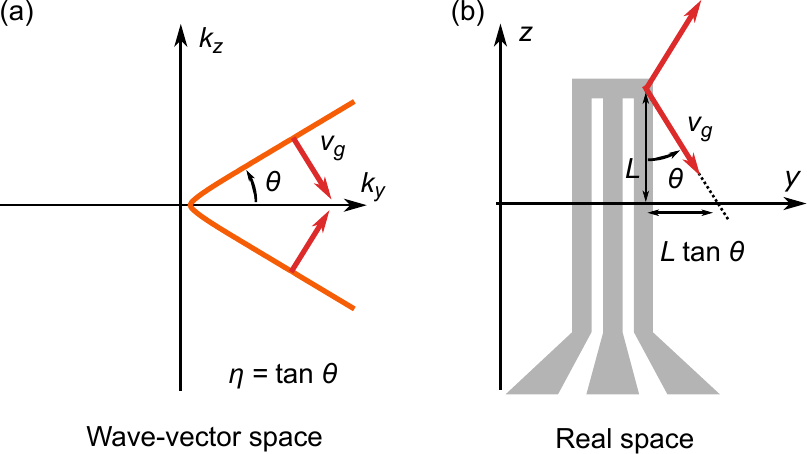}
 
\caption{Schematic picture of the caustic wave beams from the antenna corner. The group velocity is denoted as $\vector{v}_{g}$ and the asymptote of the isofrequency contours is along the angle $\theta$. (a) Wave-vector space. (b) Real space. In panel~(b), the caustic wave beam is assumed to be emitted from the corner of the antenna. The vertical distance between the antenna corner and the 1D scanning line ($z=0$) is $L$. The caustic wave beams that originate from the antenna corner arrive at $y=L \tan \theta$ on the 1D scanning line.  \label{fig:magnon_caustics}}
\end{figure}

The motion of the magnons can be captured by a wave packet picture. The wave packet propagations are perpendicular to the isofrequency contours. The isofrequency contours here have 
asymptotes defined by $\theta$:
\begin{equation}
\tan \theta \equiv \eta = \frac{k_{z}}{k_{y}}=\frac{\varphi_{z}}{\varphi_{y}}.
\end{equation}
This indicates that all magnons originate from this asymptote propagate in one direction, 
which is perpendicular to the asymptote as shown in Fig.~\ref{fig:magnon_caustics}(a). 
This leads to beam-like magnon propagations, that is, the so-called caustic wave beams~\cite{schneider2010}, 
as shown in Fig.~\ref{fig:magnon_caustics}(b). 

Let us now analyze the intense signals around the dashed lines in Figs.~\ref{fig:result1}(a) and \ref{fig:result1}(b). We state that the intense signal coming from the caustic wave beam originates from the antenna corner as shown in Fig.~\ref{fig:magnon_caustics}(b). 
Only considering the asymptotes, we can put $\cot \varphi_{x}\sur{i} \rightarrow -\jj$ in Eq.~(\ref{eq:4}). Then, Eq.~(\ref{eq:4}) becomes
\begin{multline}
 (1+\eta^2) +
 2(1+\kappa) \sqrt{1+\eta^2}\sqrt{1+\frac{\eta^2}{1+\kappa}}\\
 + (1+\kappa)^2\left(1+\frac{\eta^2}{1+\kappa}\right) -\nu^2= 0,  \label{eq:7}
\end{multline}
with $\eta = k_z/k_y$.
In Fig.~\ref{fig:magnon_caustics}(b), we show how to calculate the position
$y =L \tan \theta = L \eta $ along the 1D scanning line ($z=0$) to obtaining the estimates of the intense line shown in Fig.~\ref{fig:result1}(a), where the caustic wave beam is assumed to be emitted from the corner of the antenna. Using Eq.~(\ref{eq:7}) with $L=0.5$~mm, $M_s=166.4$~kA/m and $B\sub{ext}=\mu_{0} H\sub{ext}=109.0$~mT, $d=10.4$~$\upmu$m, and
$\Omega = 2\pi f/\omega_{M}$, the positions $y= L\eta$ as a function of frequency are depicted as the dashed blue line in Fig.~\ref{fig:result1}(a). 
The theoretical estimation nicely matches to the intense line we observed. 
On the other hand, the propagation direction of magnons generated in the inside of the antenna is restricted to \(\theta \sim \pi/2\) due to the structure of the antenna.

\begin{figure}[tbhp]
  \centering
   \includegraphics[width=8.6cm]{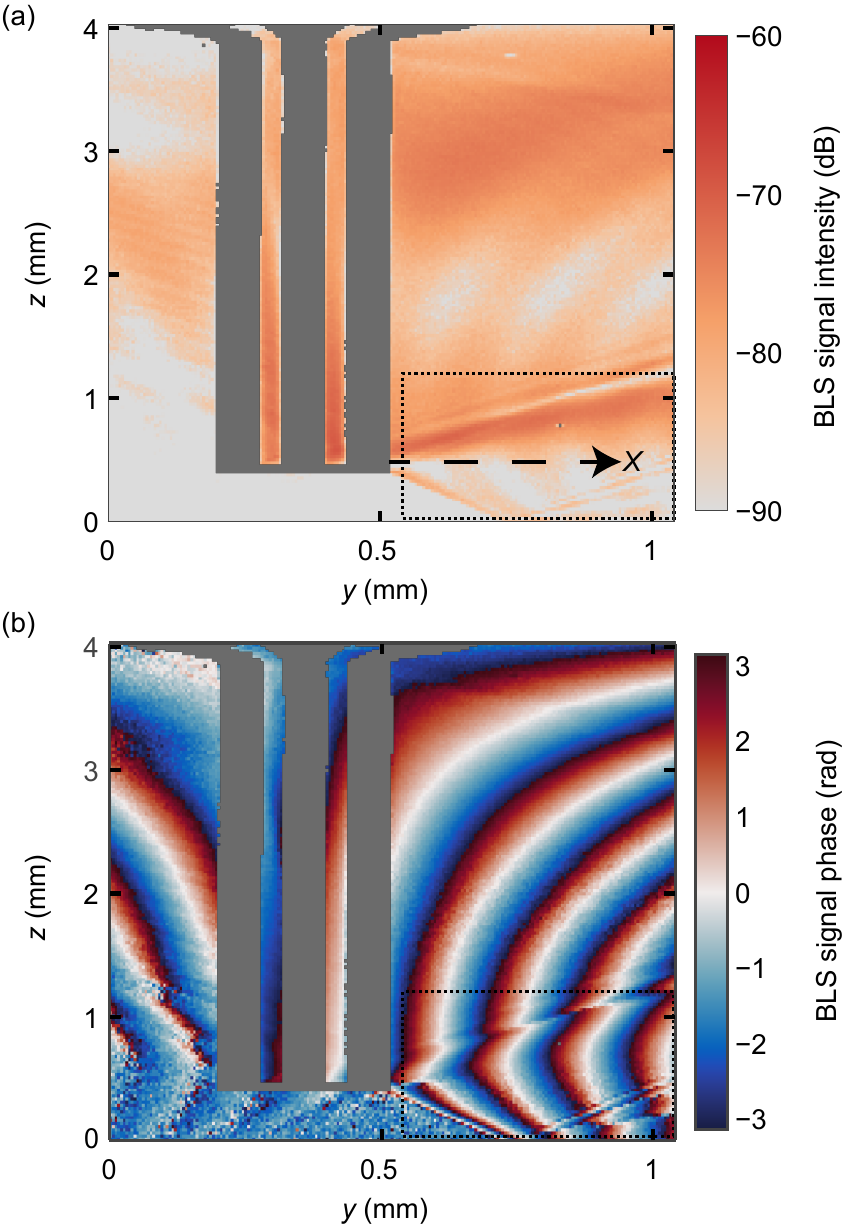}
  
 \caption{Two-dimensional real-space images of (a) the intensity and (b) the phase of the heterodyne signal from the Brillouin light scattering by magnons in the plain ferromagnetic film at the normalized microwave frequency, 
 \(\Omega = \)1.01 ($\omega /2\pi \sim 5.39$~GHz) of the excitation. 
 The dotted frame shows the area where we perform a Fourier transform for obtaining the corresponding plot in Fig.~\ref{fig:2dmap2}. 
 Note that the orientation of the antenna is different from the one shown in Fig.~\ref{fig:sample}. The aspect ratio of axes here is not unity.
 }\label{fig:2dmap1}
 \end{figure}

To experimentally address the origin of the caustic wave beam, we perform the 2D optical heterodyne imaging. 
Figures~\ref{fig:2dmap1}(a) and \ref{fig:2dmap1}(b) show the intensity and phase images of the optical heterodyne signal for magnons propagating around the corner. 
In this analysis, we use a sample different from the ones shown in Fig.~\ref{fig:sample}. 
The thickness of the YIG layer is $d=$~10.4~$\upmu$m and its antenna's length is \(\sim 3.5\)~mm~, while the length of the antenna in Fig.~\ref{fig:sample} is \(\sim 1\)~mm.
Caustic wave beams are indeed visible in Fig.~\ref{fig:2dmap1}(a). 

\begin{figure}[tbhp]
  \centering
   \includegraphics[width=8.6cm]{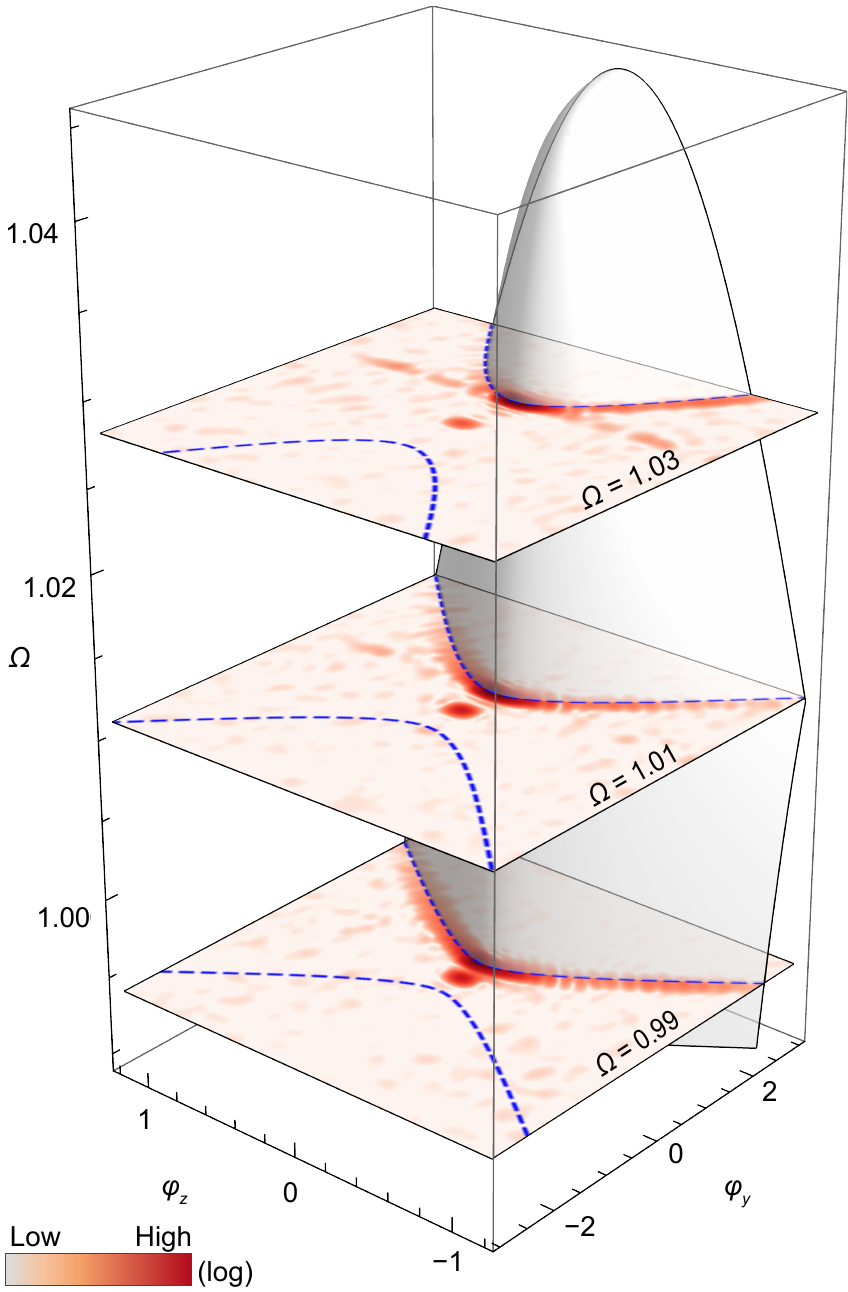}
  
 \caption{Experimentally reconstructed isofrequency contours on the planes separated by the interval $\Delta \Omega =0.02$. 
 These contours are obtained from the real-space images in Fig.~\ref{fig:2dmap1}. 
 The dashed lines and the curved surface indicate the theoretically calculated contours [Eq.~(\ref{eq:4})] with $M_s= 151$~kA/m and $B_\mathrm{ext}= 103.7$~mT.
 Note that the aspect ratio of the axes here is not unity.
 }\label{fig:2dmap2}
 \end{figure}

Next, we apply a 2D Fourier transform with Hann windows for the 2D data surrounded by the 
dotted frame in Figs.~\ref{fig:2dmap1}(a) and (b) (\(0.54 \text{~mm}< y < 1.04 \text{~mm}\), \(0.025 \text{~mm}< z < 1.2 \text{~mm}\)), and obtain 2D dispersion relations for each frequency as shown in Fig.~\ref{fig:2dmap2}. 
These results clearly indicate the hyperbolic-like nature of the isofrequency contours of Damon-Eshbach modes. The calculated isofrequency contours are also plotted. 

\bibliographystyle{apsrev4-1}
%

\end{document}